\documentclass[10pt,conference]{IEEEtran}
\usepackage{cite}
\usepackage{amsmath,amssymb,amsfonts}
\usepackage{algorithmic}
\usepackage{graphicx}
\usepackage{textcomp}
\usepackage{xcolor}
\usepackage{todonotes}
\setuptodonotes{inline}
\PassOptionsToPackage{hyphens}{url}\usepackage{hyperref}
\usepackage{siunitx}
\usepackage{amsmath}
\usepackage{flushend} 
\usepackage{tabularx}
\usepackage{diagbox}
\usepackage{enumitem}

\usetikzlibrary{shadows}
\usepackage[framemethod=tikz]{mdframed}
\newmdenv[
  leftmargin=0pt,
  rightmargin=0pt,
  usetwoside=false,
  innerleftmargin=7.5pt,
  innerrightmargin=7.5pt,
  innertopmargin=5pt,
  innerbottommargin=5pt,
  skipabove=0.5\baselineskip plus 1pt
  skipbelow=0.5\baselineskip plus 2pt 
  linecolor=black,
  backgroundcolor=white,
  linecolor=black,
  linewidth=1pt,
  roundcorner=4pt,
  nobreak=true,
  shadow=false
]{framed}

\def\BibTeX{{\rm B\kern-.05em{\sc i\kern-.025em b}\kern-.08em
    T\kern-.1667em\lower.7ex\hbox{E}\kern-.125emX}}

\newcommand{\refRQ}[1]{\hfill {\tiny $\rightarrow$\textbf{RQ#1}}}

\begin{document}

\title{Not real or too soft?\\ On the challenges of publishing interdisciplinary software engineering research 
}

\author{
    \IEEEauthorblockN{Sonja M. Hyrynsalmi\IEEEauthorrefmark{1}, Grischa Liebel\IEEEauthorrefmark{2}, Ronnie de Souza Santos\IEEEauthorrefmark{3}, Sebastian Baltes\IEEEauthorrefmark{4}}
    \IEEEauthorblockA{\IEEEauthorrefmark{1}LUT University, Finland\\
    sonja.hyrynsalmi@lut.fi} 
    \IEEEauthorblockA{\IEEEauthorrefmark{2}Reykjavik University, Iceland\\
    grischal@ru.is}
    \IEEEauthorblockA{\IEEEauthorrefmark{3}University of Calgary, Canada\\
    ronnie.desouzasantos@ucalgary.ca}
    \IEEEauthorblockA{\IEEEauthorrefmark{4}University of Bayreuth, Germany\\
    sebastian.baltes@uni-bayreuth.de} 
}


\maketitle

\begin{abstract}
The discipline of software engineering (SE) combines social and technological dimensions.
It is an interdisciplinary research field.
However, interdisciplinary research submitted to software engineering venues may not receive the same level of recognition as more traditional or technical topics such as software testing. 
For this paper, we conducted an online survey of 73 SE researchers and used a mixed-method data analysis approach to investigate their challenges and recommendations when publishing interdisciplinary research in SE.
We found that the challenges of publishing interdisciplinary research in SE can be divided into topic-related and reviewing-related challenges.
Furthermore, while our initial focus was on publishing interdisciplinary research, the impact of current reviewing practices on marginalized groups emerged from our data, as we found that marginalized groups are more likely to receive negative feedback.
In addition, we found that experienced researchers are less likely to change their research direction due to feedback they receive.
To address the identified challenges, our participants emphasize the importance of highlighting the impact and value of interdisciplinary work for SE, collaborating with experienced researchers, and establishing clearer submission guidelines and new interdisciplinary SE publication venues.
Our findings contribute to the understanding of the current state of the SE research community and how we could better support interdisciplinary research in our field.
\end{abstract}


\section{Introduction}

Software engineering (SE) is an interdisciplinary field that integrates both social and technological dimensions to enable the systematic design, development, testing, and maintenance of software systems~\cite{wang2004cognitive, melegati2021surfacing}. As a discipline subject to continuous evolution, it adapts to rapid advances in technology by employing a wide range of methodologies, standards, and tools aimed at optimizing productivity, cost-efficiency, and ensuring controllable quality throughout the software development lifecycle~\cite{boehm2006view, coltell2014finding}. Achieving these goals requires \textit{interdisciplinary} collaboration involving professionals from various domains, each contributing specialized knowledge to create reliable, scalable, and maintainable software systems~\cite{mengel1999multidisciplinary, winchester2023harmful}.

From a technical perspective, SE encompasses key tasks such as requirement elicitation, system architecture design, programming, and testing, all of which are integral to the development process~\cite{masood2022like, zulfiqar2022microtasking, valle2023soft}. On the social side, software is developed by humans, which means that it needs effective collaboration, communication, and a thorough understanding of human factors that affect team dynamics and influence the successful implementation of technical solutions~\cite{dutra2021human, dos2011action, valle2023soft}. Software engineers must understand and navigate human dynamics, ensuring that their processes and tools not only solve technical challenges but also enhance collaboration, optimize team activities, and foster innovation, thereby positively influencing the success of software projects~\cite{dutra2021human, gunatilake2024impact}.

The combination of technical expertise and human-centric approaches position SE as a truly interdisciplinary domain, where fully understanding its various aspects, characteristics, and practices requires acknowledging its multifaceted nature, which encompasses both technical processes and social dynamics~\cite{lenberg2015behavioral, wang2004cognitive, mills2011getting}. Therefore, SE research requires a comprehensive exploration of systems, practices, and tools, as well as how individuals and teams interact with these elements and each other~\cite{storey2020software}. This distinctive characteristic, where human factors and technical proficiency are equally important in the construction of software artifacts, broadens the scope of research topics and encourages the use of various different methodologies to address complex socio-technical questions in software development~\cite{glass2002research, stol2018abc, storey2020software, wohlin2015towards}.

Previous work on interdisciplinary research suggests that interdisciplinary studies often encounter resistance from traditional publication venues\cite{guimaraes2019doing}. In SE, interdisciplinary research can focus on less technical or non-traditional topics such as human factors, organizational dynamics, or socio-technical aspects. Although these topics are important for a comprehensive understanding of software development, they may not be regarded with the same level of importance as more technical subjects such as programming or testing. This can result in reviews for submissions in SE venues that are negative or even harsh toward interdisciplinary topics. Such reviews might negate the significance of interdisciplinary topics, which can cause frustration and sometimes demotivation for researchers who try to push the boundaries of knowledge by exploring different perspectives within the field~\cite{guimaraes2019doing, von2019interdisciplinary, galster2023empirical}.

Grounded in SE's interdisciplinary nature, which integrates social dynamics, human factors, and organizational behavior alongside technical components, this study seeks to explore how research that extends beyond traditional technical topics is received within the field.
Understanding this context is important because interdisciplinary research areas play a key role in expanding knowledge about the dynamics of software development. 
Therefore, our objective is to identify how publication difficulties impact research directions and opportunities by exploring the challenges researchers face in publishing and pursuing interdisciplinary SE research. Specifically, our aim is to answer the following research questions:


\begin{description}
\item[\textbf{RQ1}] Which demographics of SE researchers are associated with publishing difficulties, negative feedback, or a change of research direction?
\item[\textbf{RQ2}] What specific challenges have SE researchers faced when working on interdisciplinary topics?
\item[\textbf{RQ3}] What recommendations do participants offer for improving the research environment to better support interdisciplinary studies in SE?
\end{description}

The structure of this paper is as follows: Section 2 provides a background on interdisciplinary SE research. Section 3 outlines our methodology for conducting an online survey with SE researchers. In Section 4, we present the results of the survey. Section 5 focuses on a discussion of these findings and their implications. Finally, we conclude the study in Section 6, summarizing key insights and potential future directions.

\section{Background}

The challenges in our world have become, or have always been, complex and have even been described as `wicked' problems~\cite{lawrence2022characteristics}. Recently, in many disciplines, there has been discussion that more interdisciplinary research, which inspires or blends with other disciplines, would be needed to tackle such wicked problems. Many scholars even argue that to achieve novel findings, interdisciplinary collaboration is required~\cite{forman2005research, hessels2008re, raasch2013rise}.
For the field of SE, M{\'{e}}ndez and Passoth have argued that, especially when studying social, cultural and human-centric aspects, empirical SE should be considered an ``interdiscipline''~\cite{DBLP:journals/jss/FernandezP19}.

But what is interdisciplinary research? For example, Klein and Newell~\cite{newell1996interdisciplinary} emphasize that, although approaches may vary and debates about terminology still persist, interdisciplinary studies can be broadly understood as a method for tackling questions, solving problems, or addressing topics that are too broad or complex for a single discipline or profession to handle. Whether applied to general education, women's studies programs, or areas such as science, technology, and society, interdisciplinary studies combine insights from different fields to create a more comprehensive perspective. In this way, interdisciplinary research serves not only as a supplement but also as a means of enhancing and refining the disciplines upon which it draws. Klein and Newell also highlight that knowledge has become increasingly interdisciplinary overall. 

Terminology matters, especially when discussing terms that can have different meanings for different people, as interdisciplinary research sometimes tends to have~\cite{siedlok2014organization, huutoniemi2010analyzing, von2019interdisciplinary}. For example, research that blends with other disciplines can be called multi-, cross- or interdisciplinary research. During our research process, we utilized the definition of Klein~\cite{10.1093/oxfordhb/9780198733522.013.3} and summarized it in this way for the participants: 

\begin{quote}
\emph{``Definition of interdisciplinary software engineering research: The integration of concepts, methodology, procedures, epistemology, terminology, data, and organization of research from other disciplines into software engineering research projects.''}
\end{quote}

In general, the amount of interdisciplinary research is growing among different research fields~\cite{luke2015breaking, raasch2013rise}. However, conducting interdisciplinary research is not always easy. It requires building the right kind of collaboration opportunities, aligning various research methodologies, having open-minded funding possibilities and access to publishing platforms. Research has shown that, especially for younger academics, pressure to obtain funding could affect their willingness to engage in interdisciplinary research~\cite{luke2015breaking}.

In SE, there has been an ongoing dialogue about the importance of interdisciplinary collaboration. Similarly, other disciplines, such as Human-Computer Interaction (HCI), have also emphasized the benefits of collaborating with SE~\cite{hartson1998human}. Furthermore, in educational research, studies have shown that collaboration between engineering and design students can greatly improve the development of human-computer interaction systems, highlighting the positive impact of interdisciplinarity~\cite{kuo2019promoting}. 

Research methods constitute a large part of the interdisciplinary research discussion, and SE as a research discipline has been subjected to active empirical methods discussions for decades~\cite{seaman1999qualitative, storey2020software, lenberg2024qualitative}. As part of this discussion, it has been stated that the discipline of SE research is still a relatively young research discipline~\cite{lenberg2024qualitative}. Interdisciplinary challenges in SE may be linked to the field's young age and, in that way, the reliance on other disciplines for research methods and theoretical models. However, research has shown that when a discipline matures, research becomes more discipline-centered, influenced by a deeper understanding of the subject and academic structures~\cite{raasch2013rise}. Neighboring research areas also tend to become more similar over time, and this is especially common for STEM disciplines~\cite {mcgillivray2022investigating}. 

Previous work has shown that empirical research in SE can face different expectations from different types of reviewers.
There can be differences between whether reviewers prefer qualitative or quantitative research, different types of participants, or various uses of methodologies~\cite{galster2023empirical, bryman2007barriers}.
Interdisciplinary research in general faces challenges around publishing or being valued also in other fields than SE.
This kind of academic gatekeeping~\cite{parti2021future} can happen, especially in more `traditional' academic outlets.
One way to address this challenge has been to identify people in the community who bridge terminological and methodological gaps between disciplines~\cite{parti2021future}.

\begin{table*}
\renewcommand{\arraystretch}{1.1}
\scriptsize
\fontdimen2\font=2pt 
\centering
\caption{Online survey questionnaire.}
\begin{tabularx}{\textwidth}{| p{2.4cm} | >{\hsize=0.9\hsize}X | >{\hsize=1.1\hsize}X |}
\hline
\textbf{Variable} & \textbf{Question} & \textbf{Values and Coding} \\
\hline
\textsc{experience} \refRQ{1}
& Are you:
&	{ \renewcommand{\arraystretch}{1}
         \hspace{-8pt}
       \begin{tabular}[t]{l}
       1 = Early-stage SE researcher (up to 7 years from your first peer-reviewed paper)\\
       2 = Consolidated/middle-career SE researcher (8–12 years of research)\\
       3 = Experienced SE researcher (more than 12 years of research)
	  \end{tabular}
    }\\
\hline
\textsc{country}
& Choose the country of your affiliation from drop-down menu:
& Drop-down menu with country names\\
\hline
\textsc{gender} \refRQ{1}
& What gender do you identify as?
& Woman / Man / Non-binary / Prefer not to say / Other, please specify: (open-ended) \\
\hline
\textsc{marginalized\_group} \newline \phantom{a} \refRQ{1}
& Do you identify as belonging to a marginalised group in the context of your professional environment?
&	{ \renewcommand{\arraystretch}{1}
         \hspace{-8pt}
       \begin{tabular}[t]{l}
       0 = No \\
       1 = Yes
	\end{tabular}
    }\\
\hline
& Yes $\rightarrow$ Which group (Please select all that apply)
& Gender / Ethnicity / Age / LGBTQ+ / Other\\
\hline
\textsc{topics}
& On which interdisciplinary research topics related to software engineering do you work? Below are some examples, but feel free to list any other topics not mentioned.
& Artificial Intelligence (AI) / Cognitive Science and Psychology / Cybersecurity / Data Science and Big Data / Digital Humanities / Diversity, Equity, and Inclusion (DEI) / Education / Fairness / Health Informatics / Human-Computer Interaction (HCI) / Internet of Things (IoT) / IT Ethics / Legal Aspects of Computing / Software Business / Sustainability / Other interdisciplinary topic (open-ended)\\
\hline
\textsc{venues}
& To which software engineering conferences and journals do you usually submit your interdisciplinary software engineering research papers? Please select all that apply.
& List of publication venues (including an open-ended option for listing other venues)\\
\hline
\textsc{difficulties} \refRQ{1}
& Have you experienced difficulties related to the acceptance or review process of interdisciplinary research in software engineering?
&	{ \renewcommand{\arraystretch}{1}
         \hspace{-8pt}
       \begin{tabular}[t]{l}
       0 = No \\
       1 = Yes \\
       NA = I don't know
	\end{tabular}
    }\\ 
\hline
\refRQ{2}
& Yes $\rightarrow$ Can you provide more details on those difficulties?
& Open-ended\\
\hline
\textsc{negative\_feedback} \newline \phantom{a} \refRQ{1}
& Have you experienced negative feedback from colleagues or other researchers related to your interdisciplinary research focus?
&	{ \renewcommand{\arraystretch}{1}
         \hspace{-8pt}
       \begin{tabular}[t]{l}
       0 = No \\
       1 = Yes \\
       NA = I don't know
	\end{tabular}
    }\\ 
\hline
\refRQ{2}
& Yes $\rightarrow$ Can you provide more details on the feedback received?
& Open-ended\\
\hline
\textsc{reviews\_positive} \newline \phantom{a} 
& Would you like to share examples of some positive review comments you received about your interdisciplinary SE research? Please provide detailed comments in the box below.
& Open-ended\\
\hline
\textsc{reviews\_negative} \newline \phantom{a} \refRQ{2}
& Would you like to share examples of some negative review comments you received about your interdisciplinary SE research? Please provide detailed comments in the box below.
& Open-ended\\
\hline
\textsc{changed\_direction} \newline \phantom{a} \refRQ{1}
& Have you changed your research directions because of feedback on your interdisciplinary research?
&	{ \renewcommand{\arraystretch}{1}
         \hspace{-8pt}
       \begin{tabular}[t]{l}
       0 = No \\
       1 = Yes, due to negative feedback \\
       1 = Yes, due to positive feedback \\
       NA = I have been thinking of changing it, but haven't done so yet\\
	\end{tabular}
    }\\ 
\hline
\textsc{changes} \refRQ{3}
& What changes would you like to still see in the way interdisciplinary research is reviewed or discussed in the software engineering community?
& Open-ended\\
\hline
\textsc{advice} \refRQ{3}
& What advice would you give to researchers new to publishing interdisciplinary research in software engineering?
& Open-ended\\
\hline
\textsc{further\_comments} \newline \phantom{a} \refRQ{2,3}
& Do you have any further comments related to this survey?
& Open-ended\\
\hline
\end{tabularx}
\label{tab:questionnaire}
\end{table*}

\section{Research Process}

To answer our research questions, we conducted an anonymous online survey to collect feedback from SE researchers who have experience submitting and publishing interdisciplinary research in SE venues.
We used two different sampling strategies to recruit participants.
The objective of the first sampling strategy was to recruit researchers who have published at the International Conference on Software Engineering (ICSE) or co-located events, based on a sampling frame derived from DBLP data (\textsc{sample~1}, recruitment August-September 2024, $n=53$).
The goal of the second sampling strategy was to reach SE researchers who do not necessarily have published at ICSE or co-located events, especially more junior ones (\textsc{sample~2}, recruitment in September 2024, $n=20$).

\subsection{Sampling}

\begin{figure*}
    \centering
    \includegraphics[width=0.9\textwidth]{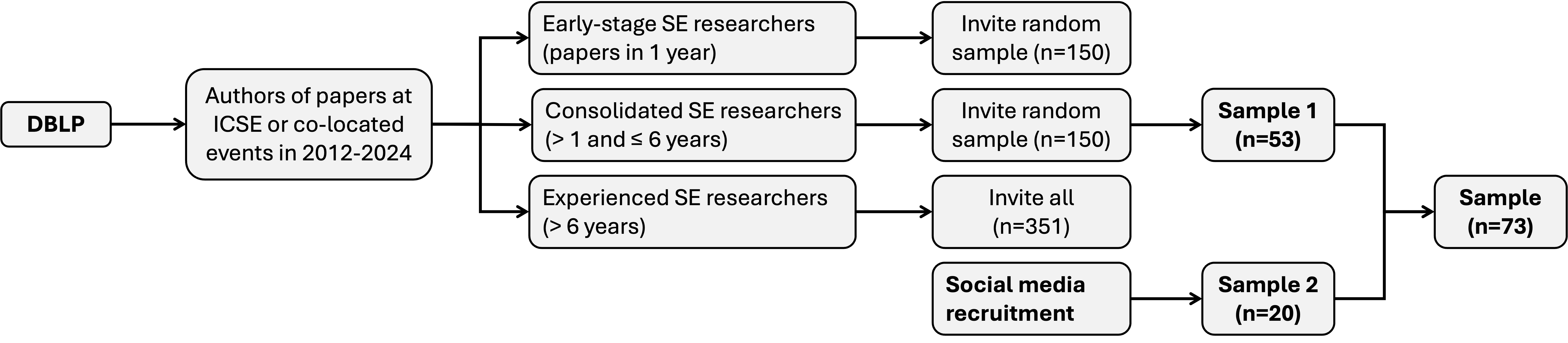}
    \caption{Overview of our sampling process.}
    \label{fig:sampling}
\end{figure*}

For \textsc{sample 1}, we decided to focus on researchers who have been publishing at ICSE or co-located events, as ICSE is the leading SE research venue, and the co-located events represent researchers from different subfields of SE.
We included authors who have published there since 2012, because the ICSE 2012 call for contributions was the first to list ``human and social aspects'' among the conference topics~\cite{ICSE2012-CfP} (based on an analysis of all ICSE calls we found online: 1997, 2001-2007, 2009-2012, 2014-2024).

We used the DBLP SPARQL interface~\cite{DBLP-SPARQL} to retrieve all authors who published in the ICSE stream~\cite{DBLP-ICSE} since 2012, including their affiliation, ORCID, and homepage link.
This resulted in \num{13082} distinct DBLP author IDs.
Drawing a random sample from this sampling frame would have resulted in many authors with one single publication in the 2012-2024 ICSE stream (\num{9506}, \qty{72.7}{\percent} of all author IDs).
Therefore, we stratified the authors, starting with one strata of very active ICSE authors.
We aimed to select the top \qty{10}{\percent} of all authors with more than one publication (\num{351} authors with $>6$ years with ICSE publications).
The second strata were the authors with only one publication, and the third strata were the remaining authors (\num{3225} authors with $>1$ and $\le6$ years with ICSE publications).
We randomly selected 150 authors from each strata, resulting in a sample of 450 authors.
For these 450 authors, we retrieved their email addresses either from their personal websites or from recent publications. 
We invited all authors with a valid email address to participate in our online survey in August 2024.
For 35 authors (\qty{7.7}{\percent}), we were unable to retrieve a valid email address.
The SPARQL query and the sampling scripts we used are part of our supplementary material~\cite{supplementarymaterial}.

We sent invitation emails at the end of August 2024.
Due to a relatively low response rate even after two reminders to participate, we later invited the remaining 201 authors from the experienced strata as well, because we noticed that experienced researchers were more likely to respond. We found email addresses for all of them, but excluded one author of this paper from the list.

\textsc{Sample~2} was collected through social media recruitment. We used social media (X, LinkedIn) to better represent the voice of less experienced researchers, whom we apparently did not reach with our systematic sampling approach for \textsc{Sample~1}. This social media sample was collected in September 2024. 
Using two different invitation links, we were able to compare this convenience sample~\cite{DBLP:journals/ese/BaltesR22} with the more systematic DBLP-based sampling.
In total, we received 73 responses, 53 recruited through systematic sampling, and 20 through social media.
Eleven of those 20 participants were early-stage researchers ($\le 7$ years from the first peer-reviewed article), versus only 7 in the first sample.


\subsection{Questionnaire Design}

The purpose of the survey was to identify the challenges that researchers face when publishing interdisciplinary SE research.
Specifically, in our questions, we focused on the following areas: demographics (academic age, gender, being a member of a marginalized group, country of affiliation), research topics and venues, experiences publishing interdisciplinary SE research, and feedback received from colleagues.
We also asked participants to provide recommendations for the community.
All survey questions are listed in Table~\ref{tab:questionnaire}.
The complete questionnaire is also part of our supplementary material~\cite{supplementarymaterial}.


We tested the questionnaire design with one experienced SE researcher ($>12$ years of research experience) and one early-stage SE researcher ($\le 7$ years since first peer-reviewed paper). Both pilot participants are not authors of this paper.

\subsection{Participants}

\begin{table*}
\centering
\caption{Participant demographics (marginalized\_group: participants could select multiple groups, $n=73$).}
\label{tab:participants}
\begin{tabularx}{0.9\textwidth}{|l|X|}
  \hline
  \textbf{Variable} & \textbf{Response Distribution} \\
  \hline
  \textsc{experience} & 18 Early-stage, 12 Mid-career, 43 Experienced \\
  \hline
  \textsc{country} & 15 USA, 8 Canada, 8 Germany, 7 Finland, 6 Italy, 3 Australia, 3 Netherlands, 3 Sweden, 18 Other, 2 NA \\
  \hline
  \textsc{gender} & 49 Man, 21 Woman, 0 Non-binary, 1 Other, 2 Prefer not to say \\
  \hline
  \textsc{marginalized\_group} & 51 No, 20 Yes (13 Gender, 8 Ethnicity, 5 LGBTQ+, 3 Age, 2 Other), 2 NA \\
  \hline
\end{tabularx}
\end{table*}

Table~\ref{tab:participants} lists the demographics of our 73 participants.
Most of the participants identified as experienced SE researchers ($43$ with $>12$ years of research experience).
The second largest group were early-stage SE researchers (18 with $\le 7$ years), followed by mid-career SE researchers ($8$ to $12$ years).
Most of the participants were from the US (15), followed by Canada (8), Germany (8), Finland (7), and Italy (6).
Only a few participants came from Asia (5) and Africa (1), and none from South America.
The majority identified as man (49), followed by woman (21), one other (Demifem), and two participants who preferred not to disclose their gender.
Twenty participants identified as belonging to a marginalized group, with gender (13), ethnicity (8), and LGBTQ+ (5) being the most frequently mentioned groups.
The most common interdisciplinary SE topics that the participants worked on were related to human-computer interaction (30), cognition/psychology (22), and education (21).
Furthermore, participants reported that their most common venues for submitting interdisciplinary SE research were ICSE (50), Springer Empirical Software Engineering (44), IEEE Transactions on Software Engineering (36), IEEE Software (32), and ACM Transactions on Software Engineering and Methodology (32).

\subsection{Data Analysis}
\label{sec:data-analysis}


For our quantitative analysis, we used \emph{Fisher's Exact Test for Count Data} as implemented in R to calculate the odds ratio and the corresponding p-value between different groups in terms of their effect on difficulties related to the acceptance or review process of interdisciplinary SE research (\textsc{difficulties}), experiencing negative feedback from colleagues or other researchers related to an interdisciplinary research focus (\textsc{negative\_feedback}), and the fact that researchers have changed their research direction based on feedback they received (\textsc{changed\_direction}).
The exact formulations of the questions can be found in Table~\ref{tab:questionnaire}.
\textsc{difficulties} and \textsc{negative\_feedback} were binary questions (yes/no) with the option to answer `I don't know.' 
For \textsc{changed\_direction}, we differentiated between `yes, due to positive feedback,' `yes, due to negative feedback,' `I have been thinking about it,` and `no.'

For our qualitative analysis, we used descriptive coding~\cite{saldana2021coding} to analyze and summarize open-ended responses from our participants.
We considered descriptive coding suitable because we wanted to make the current state of publishing interdisciplinary SE research visible, as reflected in our participants' responses. 
We then extracted all the coded text snippets, copied them into a collaborative online whiteboard tool, and further categorized them starting from the initial codes.
During the analysis, we did not distinguish by question, since the answers often overlapped, e.g., answers to difficulties in publishing interdisciplinary work in SE, negative examples of reviews, and details of negative feedback.
This analysis resulted in two high-level categories describing challenges (\textbf{RQ2}) and three high-level categories describing recommendations (\textbf{RQ3}), which are listed in Table~\ref{tab:results}.

\section{Results}

The following subsections outline our key findings based on our quantitative and qualitative data analysis. 

\subsection{Quantitative Analysis (RQ1)}

\begin{table}
\centering
\caption{Contingency table for being a member of a marginalized group and having received negative feedback.}
\label{tab:marginalized-vs-negative}
\begin{tabular}{|c|r|r|}
  \hline
  \diagbox{\scriptsize \textsc{marginalized\_group}}{\scriptsize \textsc{negative\_feedback}} & No & Yes \\
  \hline
  No & 30 & 15 \\
  \hline
  Yes & 6 & 12 \\
  \hline
\end{tabular}
\end{table}

To answer \textbf{RQ1}, we grouped the participants into two groups: experienced (43 with $> 12$ years of experience) and less experienced (30 with $\le 12$ years) researchers.
We further grouped them into man (49) and non-man (22) and into belonging to a marginalized group (20) and not belonging to such a group (51).
We then used Fisher's exact test with a significance level of \num{0.05} to determine the odds ratio between these groups for the three effects mentioned in Section~\ref{sec:data-analysis}.

In total, 47 participants reported \textsc{difficulties} in the acceptance or review process and 28 reported \textsc{negative\_feedback}.
Most of the participants (51) did not change their research direction. 
For \textsc{changed\_direction}, we compared participants who did not change their research direction (51) with those who did (17).
Of these 17, 7 changed their direction due to positive feedback and 10 due to negative feedback; 4 participants considered changing their direction.

For the effect of experience, gender, and being a member of a marginalized group on \textsc{difficulties}, we were unable to find significant differences between the groups.
For the effect on \textsc{negative\_feedback}, the differences for experience and gender were not significant, but we found that participants belonging to marginalized groups were significantly more likely to receive negative feedback than participants who did not identify themselves as belonging to such groups.
The odds ratio was $3.9$ with a \qty{95}{\percent} confidence interval of $[1.3, \infty]$ and a p-value of $0.02$.
Table~\ref{tab:marginalized-vs-negative} shows the contingency table for this relationship.

Interestingly, experience had a significant effect on \textsc{changed\_direction}, but not in the direction one would assume.
Intuitively, experienced researchers have spent more time conducting and publishing research and are therefore more likely to have changed their research direction, for example, due to feedback they received regarding their interdisciplinary research.
However, we found that the corresponding odds ratio was $0.08$ with a \qty{95}{\percent} confidence interval of $[0.0, 0.3]$ and a p-value of $<0.05 \cdot 10^{-3}$.
This means that more experienced researchers were significantly less likely to have changed their research direction.
The differences for other groups were not significant.


\begin{framed}
\textbf{RQ1 (Demographics):} Participants in marginalized groups were significantly more likely to receive negative feedback than other participants. Besides, more experienced researchers were significantly less likely to have changed their research direction.
\end{framed}

In summary, to answer \textbf{RQ1}, we found that while experience or gender did not show a significant difference, participants belonging to marginalized groups were significantly more likely to receive negative feedback compared to other participants. 
Understanding the reasons for this and identifying potential confounding factors is an important direction for future work.
This research might then motivate a more holistic role for diversity and inclusion chairs at conferences that also includes feedback on the review process.
In addition, compared to less experienced researchers, we found that experienced researchers were significantly less likely to change their research direction due to feedback they receive.
A potential explanation for this could be that less experienced researchers have a narrower view of \textsc{changed\_direction} than more experienced researchers. We discuss this aspect in Section~\ref{sec:realse}.

\subsection{Overview of Qualitative Analysis (RQ2 and RQ3)}
In our qualitative analysis, we identified two high-level categories that describe challenges researchers face when conducting and publishing interdisciplinary SE work (\textbf{RQ2}). These categories reflect the perceived challenges of the participants and the negative feedback they received related to their \textbf{research topic} and during \textbf{the review process}.
To address the reported challenges (\textbf{RQ3}), we found two categories that directly map to the two challenge categories, i.e., advice to \textbf{deal with topic-related and reviewing-related challenges}. Additionally, one more category emerged that contained general advice for researchers conducting interdisciplinary work in SE on how to \textbf{deal with the broken system}.

\begin{table}[ht]
\centering
\label{tab:results}
\caption{Results: Challenges of interdisciplinary software engineering research and participants' recommendations.}
\begin{tabular}{|m{.95\linewidth}|}

\hline
\vspace{0.25\baselineskip}
\textbf{Challenges}
\vspace{0.25\baselineskip} \\ 
\hline

\vspace{0.5\baselineskip}

\textbf{Topic-related Challenges:}
\begin{itemize}[after=\vspace{-0.5\baselineskip}]
    \item Research seen as out of scope
    \item Research seen as disconnected from SE
    \item Negative views towards an interdisciplinary research focus
    \item Research seen as irrelevant or lacking novelty
    \end{itemize} \\ 

\textbf{Reviewing-related Challenges:}  
\begin{itemize}[after=\vspace{-0.5\baselineskip}]
\item Invalid methodological criticism
\item Lack of expertise among reviewers
\item Getting interdisciplinary grants
\item Aligning interdisciplinary teams
\item Lack of professionalism and negative attitudes
\end{itemize} \\
    
\hline
\vspace{0.25\baselineskip}
\textbf{Recommendations}
\vspace{0.25\baselineskip} \\ 
\hline

\vspace{0.5\baselineskip}

\textbf{Dealing with Topic-related Challenges:} 
\begin{itemize}[after=\vspace{-0.5\baselineskip}]
\item Focus papers on SE-specific concerns
\item Choose lead author according to discipline
\item Define venue scope clearer
\item Join forces with senior researchers
\end{itemize} \\ 

\textbf{Dealing with Reviewing-related Challenges:} 
\begin{itemize}[after=\vspace{-0.5\baselineskip}]
\item Invite expert reviewers
\item Require reviewers to assess only parts of a submission
\item Clearer review criteria and standards
\item Stronger editor/chair involvement
\item Discuss expectations in interdisciplinary teams
\item Acknowledge difficulties in interdisciplinary research
\end{itemize} \\

\textbf{Dealing with a Broken System:} 
\begin{itemize}[after=\vspace{-0.5\baselineskip}]
    \item Persevere
    \item Explore other/new venues
    \item Get involved in program committees/organization positions.
    \item Keep debating the importance of interdisciplinary work
    \item Accept that interdisciplinary work requires more effort and time
    \item Avoid doing interdisciplinary research in SE
\end{itemize} \\

\hline
\end{tabular}
\end{table}

\subsection{Topic-related Challenges (RQ2)} 
Topic-related challenges describe participants' experiences of how their research approaches were not seen as a part of SE.
That is, many participants shared experiences of colleagues or reviewers considering their work out of scope:
\vspace{2pt}
\begin{quote}
    \textit{"Mainly from CS/SE, who would argue that this is not CS/SE and should be targeted at psychology, vocational studies, etc. venues."} - ID37
\end{quote}
\vspace{2pt}
\begin{quote}
\textit{"Just some regular commenting [...] is my research in the scope of SE or IS [Information Systems]?"}
ID70
\end{quote}
\vspace{2pt}

Similarly, several participants shared difficulties in getting colleagues to understand why a topic has connections to SE.
\vspace{2pt}
\begin{quote}
\textit{"Some reviewers failed to see the implications of legal aspects in software engineering."} 
ID7
\end{quote}
\vspace{2pt}

Beyond questioning the scope of research work, several participants shared outright negative views expressed by their colleagues, often directly aimed at the respective person's research profile.

\vspace{2pt}
\begin{quote}
\textit{``'I don’t see a CS syllabus coming from you'. Luckily those were minority and I've got 4 job offers.''}
ID73
\end{quote}
\vspace{2pt}
\begin{quote}
\textit{``They were joking about my super-soft perspective on software engineering, not taking it really seriously.''}
ID28
\end{quote}
\vspace{2pt}

In a slightly different direction, participants shared views that their work was perceived as lacking novelty or being not relevant.

\vspace{2pt}
\begin{quote}
\textit{``Basically along the lines of what's the point.''}
ID63
\end{quote}
\vspace{2pt}
\begin{quote}
\textit{``Reviewers from SE stating that the results are not surprising and they have known that all along ''}
ID64
\end{quote}
\vspace{2pt}

\subsection{Reviewing-related Challenges (RQ2)} 
Challenges in this category focused on the feedback received during the publication process. Several participants highlighted that the review process itself posed significant barriers to publishing interdisciplinary research in SE. These challenges often led to frustration, as participants felt that their work was unfairly judged:

\vspace{2pt}
\begin{quote}
   \textit{``He or she may start with a small number of interviews. But for marginalized groups, it is quite hard to get interviews, and for qualitative data, you do not need hundreds of interviews.''}
   ID67
\end{quote}
\vspace{2pt}

This challenge was paired with several experiences of potentially invalid criticism in reviews related to methodology.

\vspace{2pt}
\begin{quote}
   \textit{``Too few respondents in expert interview.''}
   ID1
\end{quote}
\vspace{2pt}
\begin{quote}
   \textit{``The most recent example was someone complaining that the two factors in a factorial experiment weren't 'equivalent', which only indicates that the reviewer doesn't know what a factorial experiment is.''}
   ID32  
\end{quote}
\vspace{2pt}

The most common experience in this category is related to a lack of expertise among reviewers.
In total, 14 participants shared examples of a lack of reviewer expertise, both related to methodology and the interdisciplinary nature of their topic.

\vspace{2pt}
\begin{quote}
   \textit{``Usually the reviewers are not familiar with the other relevant domains. Even ML familiarity is hard to find.''}
   ID55  
\end{quote}
\vspace{2pt}
\begin{quote}
   \textit{``Reviewers often don't have the expertise to properly judge the multi-disciplinary work. For example, it is discouraging when parts of a paper that report research led by a psychologist or AI researcher are criticized by an SE reviewer for being incorrect [..].''}
   ID3
\end{quote}
\vspace{2pt}

Although our survey focused on publishing interdisciplinary research, some participants also highlighted the challenges of interdisciplinary research prior to publication.
In particular, they mentioned the grant review process and that it is challenging to satisfy an interdisciplinary review panel.

\begin{quote}
   \textit{``Grant proposals and getting them accepted. Because you typically get one reviewer from each main discipline (say, psychology and SE), and you can fulfill neither requirements completely.''}
   ID64
\end{quote}

In contrast to a lack of reviewer expertise, participants reported challenges aligning different expectations in different disciplines. These expectations could range from the way of writing or structuring a paper, the terminology used, to methodological standards.

\vspace{2pt}
\begin{quote}
    \textit{``Sometimes we (from CS) are not used to the other areas' theoretical frameworks, instruments, and styles. [..] It is always a learning experience, but it is hard to adapt to so many variables.''}  ID37  
\end{quote}
\vspace{2pt}
\begin{quote}
    \textit{``Reviewers often may have very different disciplinary backgrounds, and will ask for different and even occasionally conflicting revisions.''}
    ID50  
\end{quote}
\vspace{2pt}

Finally, we received various comments related to a lack of professionalism among reviewers or a hostile culture toward certain topics or types of research.
For instance, several participants mentioned that reviewers' comments were based on personal beliefs and were nonconstructive.
\vspace{2pt}
\begin{quote}
    \textit{``Criticizing design decisions on operationalization based on personal preference.''}
    ID64
\end{quote}
\vspace{2pt}
\begin{quote}
    \textit{``...'the actual question you should have addressed in that study was [something the person prefers]'..."}
    ID64
\end{quote}
\vspace{2pt}

\subsection{Summary of Challenges (RQ2)}

To answer \textbf{RQ2}, our qualitative analysis revealed two main challenges when working with interdisciplinary SE research. For topic-related challenges, participants reported that their research was often seen as out of scope or not relevant to the field of SE. For reviewing-related challenges, participants mentioned frequent experiences of invalid review criticism, lack of reviewer expertise, or even lack of professionalism among reviewers. 

\begin{framed}
\textbf{RQ2 (Challenges):} Topic-related challenges include interdisciplinary research being considered out of scope or not relevant to the field of SE. Reviewing-related challenges include invalid review criticism, lack of reviewer expertise, or even lack of professionalism among reviewers.
\end{framed}

\subsection{Dealing with Topic-related Challenges (RQ3)}
To deal with topic-related challenges, participants recommend writing papers with a narrow focus so that it is clear what the impact on SE is. 
\vspace{2pt}
\begin{quote}
    \textit{``Focus on the impact to SE''}
    ID48  
\end{quote}
\vspace{2pt}

If the research team is interdisciplinary, the lead author should be the one who has expertise in the field of the venue the paper is submitted to.
\vspace{2pt}
\begin{quote}
    \textit{``For each paper, decide on the angle - is it a SE paper or an AI paper, health paper, etc. [..]. Based on the angle, the lead author should be the expert in the discipline of the angle.''}
ID3  
\end{quote}
\vspace{2pt}

However, conferences and journals should also clearly define their scope and allow for interdisciplinary submissions.

\vspace{2pt}
\begin{quote}
    \textit{``Conferences and journals should better define their boundaries.''}
    ID7  
\end{quote}
\vspace{2pt}
\begin{quote}
    \textit{``I would like to see in 'Submission guidelines' or 'For authors' part of journals website a type of submission related to interdisciplinary''}
ID41  
\end{quote}
\vspace{2pt}

As advice to other researchers, our participants suggested joining forces with strong researchers or senior researchers with a known track record, as this can help get work accepted.
In a slightly more cynical direction, a few participants suggested that becoming an established researcher before publishing interdisciplinary research helps.

\vspace{2pt}
\begin{quote}
    \textit{``Have a strong publication record or be an exceedingly good writer''}
    ID46  
    \end{quote}
    \vspace{2pt}
\begin{quote}
    \textit{``To team up with a more senior researcher who stands a better chance of being listened to''}
    ID61  
    \end{quote}
    \vspace{2pt}
\begin{quote}
    \textit{``Have a good supervisor with an established name, because then there is a good chance to make it as researcher.''}
    ID64  
    \end{quote}
\vspace{2pt}

\subsection{Dealing with Reviewing-related Challenges (RQ3)}

Among our participants' advice to deal with reviewing-related challenges, by far the most frequently mentioned aspect was that expert reviewers need to be invited, potentially from other disciplines where appropriate.
In addition to this point, one participant raised the possibility that not all reviewers need to review/comment on the entire manuscript, but only the parts they are experts in.

\vspace{2pt}
\begin{quote}
    \textit{``Especially with journals, editors can try to create a balanced set of reviewers, i.e., reviewers that cover each of the subtopics.''}
    ID30  
     \end{quote}
     \vspace{2pt}
\begin{quote}
    \textit{``A review model that is open to invite new reviewers in the case the PC pool is not an expert on the topic.''}
    ID5  
     \end{quote}
     \vspace{2pt}
\begin{quote}
    \textit{``It should be accepted that a paper may need multiple reviewers each speaking to their expertise and not expected to review the sections that they lack experience with.''}
    ID63  
     \end{quote}
\vspace{2pt}

On the reviewer and editor side, participants suggested clearer reviewer criteria and stricter adherence to existing standards, such as the ACM SIGSOFT Empirical Standards~\cite{empiricalstandards}. 

\vspace{2pt}
\begin{quote}
    \textit{``Reviewers should assess papers against published evidence standards like the SIGSOFT ones. Assessing a paper by writing an essay is stupid.''} 
    ID32  
\end{quote}
\vspace{2pt}
\begin{quote}
    \textit{``Making criteria more explicit, and make a discussion about them transparent.''} 
    ID1  
\end{quote}
\vspace{2pt}

This also included comments that editors need to play a stronger role and handle reviews that do not adhere to existing standards or are outright unprofessional.

\vspace{2pt}
\begin{quote}
    \textit{``I would love to see editors/chairs push back or even discount reviews that are just stubborn.''} 
    ID54  
\end{quote}
\vspace{2pt}
\begin{quote}
    \textit{``I also believe that editors are not exercising thoughtfulness about their role in this process, and that this community needs to take seriously the covert replication crisis that is undoubtedly plaguing it.''}
    ID57  
\end{quote}
\vspace{2pt}

Finally, several comments related to how the author team writes up their research.
This included discussing expectations and publication practices early on and aligning on a common way of writing.

\vspace{2pt}
\begin{quote}
    \textit{``Have good cooperations with researchers that have already conducted interdisciplinary research, and that know how to set up, e.g., a user study. Otherwise, you will spend a lot of work on a study that has a fundamental flaw.''}
    ID26  
    \end{quote}
    \vspace{2pt}
\begin{quote}
    \textit{``Find likeminded interdisciplinary researchers and share your experiences and publication strategies!''}
    ID4  
    \end{quote}
    \vspace{2pt}
    
Furthermore, several comments related to the inherent difficulty of conducting interdisciplinary research, advising to be prepared to learn, but also to acknowledge that interdisciplinary research is difficult.

\vspace{2pt}
\begin{quote}
    \textit{``Acknowledging the increased effort required to achieve interdisciplinary research.''}
    ID1  
    \end{quote}
    \vspace{2pt}
\begin{quote}
    \textit{``Lean into theories in the associated discipline, and adapt them, and bring them into our field. To me, SE papers are mostly disconnected, isolated, singular findings that don't really add up to broader knowledge.''}
    ID27  
    \end{quote}
\vspace{2pt}

\subsection{Dealing with a Broken System (RQ3)}
As a final category of advice, we received many recommendations related to conducting and publishing interdisciplinary SE research in the current system, that is, under the various difficulties we described earlier.

The most common suggestion was the importance of perseverance and ``not giving up''.
Participants encouraged researchers to keep trying when faced with rejections and discouraging comments.
The wording \textit{"Do not give up"} occurred many times. As one participant advised:

\vspace{2pt}
\begin{quote}
    \textit{``Stay strong and believe in your research! Make sure that your methodology section is crystal clear and strong - then they can't do anything.''}
    ID23  
\end{quote}

Frequent comments are also related to being ready to have long discussions with reviewers, to educate them about the methodology, or to convince them of the relevance of the topic.

\begin{quote}
    \textit{``Educate the reviewer in the paper.''}
    ID47  
\end{quote}
\begin{quote}
    \textit{``Be prepared to have long anonymous discussing with referees when responding to their feedback!''}
    ID9  
\end{quote}
\vspace{2pt}

An alternative suggestion raised was to consider alternative venues or create new venues that appreciate an interdisciplinary focus:

\vspace{2pt}
\begin{quote}
    \textit{``Do not publish in software engineering venues. Rather, publish in the venues of your other fields which have better established empirical, methodological, and scientific norms.''}
    ID57  
\end{quote}    
\begin{quote}
    \textit{``Create new publications venues.''}
    ID14  
\end{quote}
\vspace{2pt}

In terms of long-term changes in the SE community, some participants suggested that we need to continue debating the observed issues in the community, participate in program committees and organizational roles, and change policies and attitudes over time.

\vspace{2pt}
\begin{quote}
    \textit{``Get involved in program committees and editorial boards to influence others.''}
    ID61  
\end{quote}
\begin{quote}
    \textit{``Discussion, discussion about the challenges!''}
    ID23  
\end{quote}
\vspace{2pt}

In the same direction, we also received many responses to the open-ended \textsc{further\_comments} question (see Table~\ref{tab:questionnaire}) expressing interest and appreciation in this study topic, and noting that the discussion of challenges in publishing interdisciplinary SE research needs to continue.

Finally, it is important to acknowledge that we even received comments that recommended other researchers to outright avoid (trying to) publish interdisciplinary SE research.

\vspace{2pt}
\begin{quote}
    \textit{``Honestly, it is better to avoid it.''}
    ID55  
    \end{quote}
\begin{quote}
    \textit{``Unless you are passionate about it, don't do it. There's so much of an uphill battle to get it published, it is not worth it unless you are fascinated by the topic.''}
    ID63  
    \end{quote}
\vspace{2pt}

\subsection{Summary of Recommendations (RQ3)}

For \textbf{RQ3}, participants offered various recommendations on how to deal with the challenges mentioned for interdisciplinary research in SE.
These recommendations included writing papers with a narrow focus on specific SE concerns, joining forces with senior researchers, or inviting expert reviewers.
Some participants even suggested to avoid conducting interdisciplinary SE research altogether.
Moreover, several comments indicated the need for a continued debate on publication challenges to change the community over time, for example, by revising submission guidelines or creating new venues for interdisciplinary SE research.

\begin{framed}
\textbf{RQ3 (Recommendations):} Recommendations that our participants mentioned include focusing papers narrowly on specific SE concerns, but also including senior researchers in the team or establishing clearer submission guidelines or new venues for interdisciplinary SE research. Participants also expressed the need for a continuous debate about the identified challenges. 
\end{framed}

\section{Discussion}

Building on the idea of Klein and Newell~\cite{newell1996interdisciplinary} that interdisciplinary research serves not only as a supplement but also as a means of enhancing and refining the disciplines it draws upon, our research focused on exploring the state of the art of interdisciplinary publishing in SE. Through our survey, our goal was to identify the challenges researchers face and to gather insights that can help improve the SE community's approach to interdisciplinary collaboration and publication.

\subsection{Is Real Software Engineering Interdisciplinary?}
\label{sec:realse}

Our findings suggest that SE, as a young research discipline, is still maturing in its acceptance of interdisciplinary research. This was especially visible in our participants' experiences of challenges and negative feedback.

Interestingly, more mature researchers are less likely to change their research directions, although they are receive frequent negative feedback. This could indicate that they are more used to shifting their research focus over time and may not perceive these shifts as `changing directions' in the same way as early-career researchers do.

People in the community matter. Although many of our participants highlighted challenges related to the scope, fit of their work, and reviewer feedback, the responses also emphasized the importance of colleagues. On the one hand, we received anecdotal evidence of colleagues showing negative views of interdisciplinary SE research and belittling work as, e.g., `too soft'. On the other hand, many of the suggestions relate to teaming up with the `right' people, referring both to expertise/experience and to their attitude. This has an interface to the discussion on what constitutes legitimate SE research. 

Although participants suggested several ways to address the topic- or reviewing-related challenges, such as creating new venues for interdisciplinary SE research or focusing on educating reviewers, there were signs that not all of the challenges could be addressed by the authors' own efforts. For example, we found several signs of a lack of professionalism, such as hostile review comments, potentially invalid criticism, and personal opinions. These are signs of a broken system and unprofessionalism in the community overall. 
These issues relate to different types of reviewers~\cite{galster2023empirical} and also to different perspectives and attitudes towards empirical SE research~\cite{seaman1999qualitative, dybaa2011qualitative, galster2023empirical}. Our participants highlighted that the variation in these attitudes often challenges interdisciplinary research. Future work should therefore investigate to what extent these reported challenges relate to the use of empirical research methods, especially qualitative methods, and to what extent the challenges relate more to the interdisciplinary nature of the research.    

As we further discuss in Section~\ref{sec:threats}, the term `interdisciplinary research' is not well-defined~\cite{siedlok2014organization, huutoniemi2010analyzing}, particularly in SE. For example, the call for papers in the ICSE 2025 Software Engineering in Society track (SEIS)~\cite{ICSE-SEIS-2025} lists interdisciplinary research under 'Software Engineering for Sciences, Design, Arts and Engineering'. Arguably, various other points in the same category and other categories qualify as interdisciplinary as well, e.g., medicine and public health or physical sciences. 

To foster a more open and supportive research environment, especially for new ideas and approaches, the SE community must work collectively toward making interdisciplinary research more visible and valued. One specific point made by Parti and Szigeti~\cite{parti2021future} discusses the need for `liaison officers' or `interpreters' who can bridge terminological and methodological gaps between different disciplines. We believe that these could be the people who already work on interdisciplinary research in SE but have a technical SE background.

A final point of discussion is the question what makes SE special, or to what extent our challenges differ from interdisciplinary research in other fields.
In this direction, we believe the socio-technical nature of the SE research discipline and the broad diversity of empirical methods used in the field~\cite{seaman1999qualitative} are also visible in our data. This diversity of methods potentially explains a general lack of reviewer expertise, as well as resistance to some of the methods due to a lack of understanding. In other fields, there might be a much smaller subset of `standard' methods that are applied.
Similarly, computers and software are increasingly being used in all aspects of science and society.
While this can lead to groundbreaking results, as very recently evidenced by the 2024 Nobel Prizes in Physics and Chemistry being awarded to Computer Scientists, it also raises the question to what extent SE practices differ in other fields.
Finally, applying SE across a variety of technical and non-technical domains also changes the socio-technical context, as SE is applied in a broader social context and with human stakeholders of varied backgrounds.

Our aim in this research was to explore the current state of the art of interdisciplinary SE research. We see this study as a starting point for a broader discussion of the importance, challenges, and possibilities of interdisciplinary research in SE.
In the future, it would be beneficial to build on our findings by conducting more in-depth interviews or case studies investigating the challenges we have identified.
We believe that a continued debate on the usefulness and need to study interdisciplinary topics in SE is required.

\subsection{Threats to Validity}
\label{sec:threats}
Following Wohlin et al.~\cite{DBLP:books/sp/WohlinRHORW24}, we discuss threats to the construct, external, and internal validity of our study.
We do not discuss conclusion validity as we do not draw causal conclusions from the variables we analyzed.

\subsubsection{Construct Validity}
The construct validity refers the degree to which the scales and metrics used actually measure the intended properties~\cite{DBLP:conf/ease/RalphT18}.
In general, we discussed the questionnaire repeatedly with respect to the length and clarity of questions.
Furthermore, we piloted the survey with two researchers, one experienced and one junior.
However, the formulation `interdisciplinary research', which we use in several questions, is subject to our participants' understanding of what constitutes interdisciplinary research in SE.
The disciplinary boundaries of SE are blurry, as various comments in our survey show.
Therefore, participants might have had different interpretations of this concept.
To address this threat, we provided a definition and some examples in our survey and social media materials.
We created the list of interdisciplinary research topics in our questionnaire ourselves.
To mitigate the threat of it being incomplete, we added an open-ended answer option.
Moreover, the corresponding question is only supplemental to the results presented in this paper.


\subsubsection{External Validity}
The external validity refers to the degree to which our results generalize to other contexts~\cite{DBLP:journals/ese/BaltesR22}.
First, our focus is SE research, and we do not claim that our results generalize to other disciplines.
Then, the question is whether our samples, one focused on ICSE and co-located events and one derived using social media recruitment, sufficiently capture the SE discipline.
Since ICSE is the leading SE conference and we included all co-located events and all proceedings since 2012, our sample is rather broad and we assume that it is a good approximation of the breadth of our field.
There could be a survivorship bias present in our samples, as many researchers may have left research or the SE field out of frustration over getting their work rejected or not being appreciated by colleagues.
However, it is likely that the authors we have reached with our ICSE sample have previously had papers rejected.
The inclusion of co-located events and the addition of our social media sample mitigate this threat.
Another threat could be that conferences are less geographically inclusive than journals, because authors have to be able to afford conference attendance to present their work.
This threat is visible in the fact that we did not manage to attract responses from South America and only a few responses from Asia and Africa. This limits the generalizability of our findings to the global SE research community.
This threat could be addressed by re-running our study with journal authors.
In terms of the experience of our participants, we first struggled to attract responses from early-career researchers.
We addressed this issue by launching a second convenience sampling round on social media, which attracted more junior SE researchers.
Finally, both of our samples probably have a selection bias, as researchers who experienced more negative feedback are more likely to respond to our survey, thus potentially skewing the findings.

\subsubsection{Internal Validity}
The internal validity refers to the degree to which we can rule out alternative explanations for our results~\cite{brewer2000research}.
In terms of our quantitative finding that participants belonging to marginalized groups were significantly more likely to receive negative feedback than participants who did not identify themselves as belonging to marginalized groups, there might be confounding factors that explain the negative experiences of that group.
Future research could focus on this particular subgroup of SE researchers to better understand their specific challenges.
We do not know if and how reviewers were subtly biased when reviewing papers submitted by authors from identifiable marginalized groups based on author information in the papers.
In this context, it should be noted that ICSE has been using double-blind reviews since 2018.
However, we did reach out to authors of ICSE papers prior to 2018 and, moreover, not all co-located events have moved to double-blind reviews.
In terms of our finding that more experienced researchers were significantly less likely to have changed their research direction, there can of course be different explanations for this.
For example, those researchers might have been particularly successful in attracting funding for traditional SE research topics, and hence there was no need to change directions.

\section{Conclusion}
In this paper, we highlight the interdisciplinary nature of software engineering (SE) research.
We present findings from a survey with SE researchers from diverse backgrounds conducting interdisciplinary SE research.
Based on their answers, we discuss challenges and recommendations for interdisciplinary research in SE. 

Our results reveal that there are currently various challenges in publishing interdisciplinary SE research related to research topics and the review process (see Table~\ref{tab:results}).
In addition to identifying challenges, our participants also offered recommendations on how to deal with the identified challenges.
Although experience or gender did not show a significant difference, we found that participants belonging to marginalized groups were significantly more likely to receive negative feedback compared to other participants. Furthermore, experienced researchers were significantly less likely to change their research direction due to feedback compared to less experienced ones.

Although participants offered several ways to address the topic- and reviewing-related challenges, such as educating reviewers, revising submission guidelines, or creating new publication venues for interdisciplinary SE research, there were signs of a lack of professionalism, hostile comments, and invalid criticism, sometimes arising from too opinionated reviews. Addressing these challenges requires continued long-term and community-wide efforts to better define and support interdisciplinary SE research.

\section*{Acknowledgments}

We express our deepest gratitude to all participants who took the time to share their experiences and insights in our online survey.
We also thank all the colleagues who shared our study on social media, those who participated in the pilot phase of the survey, and those who provided feedback and encouragement during data collection.
We truly value your support and input. 


\bibliographystyle{ieeetr}

\bibliography{ref}

\end{document}